\documentclass[11pt,a4paper,reqno,intlimits,sumlimits]{amsart}

\usepackage{amssymb}
\usepackage{chicago}
\usepackage{color}
\usepackage[mathscr]{euscript}
\usepackage{xspace}
\usepackage{epsfig,rotating}
\usepackage{enumerate}

\DeclareMathAlphabet{\mathpzc}{OT1}{pzc}{m}{it}

\usepackage[bookmarksopen,pdfstartview=FitH]{hyperref}
\hypersetup{colorlinks,%
            citecolor=black,%
            filecolor=black,%
            linkcolor=black,%
            urlcolor=black}%

\begin{document}

\frenchspacing

\theoremstyle{plain}
\newtheorem{theorem}{Theorem}[section]
\newtheorem{lemma}[theorem]{Lemma}
\newtheorem{proposition}[theorem]{Proposition}
\newtheorem{corollary}[theorem]{Corollary}

\theoremstyle{definition}
\newtheorem{remark}[theorem]{Remark}
\newtheorem{definition}[theorem]{Definition}
\newtheorem{assumption}{Assumption}
\newtheorem*{assuL}{Assumption ($\mathbb{L}$)}
\newtheorem*{assuAC}{Assumption ($\mathbb{AC}$)}
\newtheorem*{assuEM}{Assumption ($\mathbb{EM}$)}
\renewcommand{\theequation}{\thesection.\arabic{equation}}
\numberwithin{equation}{section}

\renewcommand{\thetable}{\thesection.\arabic{table}}
\numberwithin{table}{section}

\renewcommand{\thefigure}{\thesection.\arabic{figure}}
\numberwithin{figure}{section}

\newcommand{\Law}{\ensuremath{\mathop{\mathrm{Law}}}}
\newcommand{\loc}{{\mathrm{loc}}}
\newcommand{\Log}{\ensuremath{\mathop{\mathcal{L}\mathrm{og}}}}

\let\SETMINUS\setminus
\renewcommand{\setminus}{\backslash}

\def\stackrelboth#1#2#3{\mathrel{\mathop{#2}\limits^{#1}_{#3}}}

\newcommand{\prozess}[1][L]{{\ensuremath{#1=(#1_t)_{0\le t\le T_*}}}\xspace}
\newcommand{\prazess}[1][L]{{\ensuremath{#1=(#1_t)_{0\le t\le T_*}}}\xspace}
\newcommand{\pt}[1][]{{\ensuremath{\P_{T_{#1}}}}\xspace}
\newcommand{\ts}[1][]{\ensuremath{T_{#1}}\xspace}
\def\P{\ensuremath{\mathrm{I\kern-.2em P}}}
\def\E{\mathrm{I\kern-.2em E}}

\def\bF{\mathbf{F}}
\def\F{\ensuremath{\mathcal{F}}}
\def\R{\ensuremath{\mathbb{R}}}
\def\C{\ensuremath{\mathbb{C}}}
\def\bt{\ensuremath{\mathbf{T}}}

\def\Rmz{\R\setminus\{0\}}
\def\Rdmz{\R^d\setminus\{0\}}
\def\Rnmz{\R^n\setminus\{0\}}

\def\Rp{\mathbb{R}_{\geqslant0}}

\def\lev{L\'{e}vy\xspace}
\def\lib{LIBOR\xspace}
\def\lk{L\'{e}vy--Khintchine\xspace}
\def\smmg{semimartingale\xspace}
\def\smmgs{semimartingales\xspace}
\def\mg{martingale\xspace}
\def\tih{time-inhomoge\-neous\xspace}

\def\eqlaw{\ensuremath{\stackrel{\mathrrefersm{d}}{=}}}

\def\ud{\ensuremath{\mathrm{d}}}
\def\dt{\ud t}
\def\ds{\ud s}
\def\dx{\ud x}
\def\dy{\ud y}
\def\dsdx{\ensuremath{(\ud s, \ud x)}}
\def\dtdx{\ensuremath{(\ud t, \ud x)}}

\def\intrr{\ensuremath{\int_{\R}}}

\def\MM{\ensuremath{\mathscr{M}}}
\def\ME{\mathbb{M}}

\def\EM{\ensuremath{(\mathbb{EM})}\xspace}
\def\ES{\ensuremath{(\mathbb{ES})}\xspace}
\def\AC{\ensuremath{(\mathbb{AC})}\xspace}
\def\LL{\ensuremath{(\mathbb{L})}\xspace}

\def\ott{{0\leq t\leq T_*}}

\def\e{\mathrm{e}}
\def\eps{\epsilon}

\def\half{\frac{1}{2}}
\def\LibT{L(t,T_i)}
\def\MeaT{\P_{T_{i+1}}}
\def\volT{\lambda(s,T_i)}
\def\vol2T{\lambda^2(s,T_i)}
\def\LevT{H_s^{T_{i+1}}}

\title{Numerical Methods for the L\'evy LIBOR model}

\author[A. Papapantoleon]{Antonis Papapantoleon}
\author[D. Skovmand]{David Skovmand}

\address{Institute of Mathematics, TU Berlin, Stra\ss e des 17. Juni 136,
         10623 Berlin, Germany \& Quantitative Products Laboratory,
         Deutsche Bank AG, Alexanderstr. 5, 10178 Berlin, Germany}
\email{papapan@math.tu-berlin.de}

\address{Aarhus School of Business, Aarhus University, Fuglesangs All\'e 4,
         8210 Aarhus V, Denmark}
\email{davids@asb.dk}

\
\date{\today}\pagestyle{myheadings}
\maketitle

\begin{abstract}
The aim of this work is to provide fast and accurate approximation schemes
for the Monte-Carlo pricing of derivatives in the \lev \lib model of
\citeN{EberleinOezkan05}. Standard methods can be applied to solve the
stochastic differential equations of the successive \lib rates but the
methods are generally slow. We propose an alternative approximation
scheme based on Picard iterations. Our approach is similar in accuracy to
the full numerical solution, but with the feature that each rate is, unlike
the standard method, evolved independently of the other rates in the term
structure. This enables simultaneous calculation of derivative prices of
different maturities using parallel computing. We include numerical illustrations
of the accuracy and speed of our method pricing caplets.
\end{abstract}

\section{Introduction}
\label{intro}

 The \lib market model has become a standard model for the pricing of interest
 rate derivatives in recent years. The main advantage of the LIBOR model in
 comparison to other approaches, is that the evolution of discretely compounded,
 market-observable forward rates is modeled directly and not deduced from the
 evolution of unobservable factors. Moreover, the log-normal \lib model is
 consistent with the market practice of pricing caps according to Black's formula
 (cf. \citeNP{Black76}). However, despite its apparent popularity, the \lib
 market model has certain well-known pitfalls.

 On the one hand, the log-normal \lib model is driven by a
 Brownian motion, hence it cannot be calibrated adequately to the observed market
 data. An interest rate model is typically calibrated to the implied volatility
 surface from the cap market and the correlation structure of at-the-money
 swaptions. Several extensions of the \lib model have been proposed in the
 literature using jump-diffusions, \lev processes or general semimartingales as
 the driving motion (cf. \citeNP{GlassermanKou03}, \citeANP{EberleinOezkan05} 
 \citeyearNP{EberleinOezkan05}, \citeNP{Jamshidian99}), or incorporating
 stochastic volatility effects (cf. e.g. \citeNP{AndersenBrothertonRatcliffe05}).

 On the other hand, the dynamics of LIBOR rates are not tractable under
 every forward measure due to the random terms that enter the dynamics
 of \lib rates during the construction of the model. In particular, when the
 driving process has continuous paths the dynamics of LIBOR rates are tractable
 under their corresponding forward measure, but they are not tractable under any
 other forward measure. When the driving process is a general semimartingale,
 then the dynamics of \lib rates are not even tractable under their very own
 forward measure. Consequently: if the driving process is a \textit{continuous} \smmg caplets can be
       priced in closed form, but \textit{not} swaptions or other multi-LIBOR
       derivatives.
 However, if the driving process  is a \textit{general} \smmg, then even caplets
       \textit{cannot} be priced in closed form.
 The standard remedy to this problem is the so-called ``frozen drift''
 approximation, where one replaces the random terms in the dynamics of \lib rates
 by their deterministic initial values; it was first proposed by
 \shortciteN{BraceGatarekMusiela97} for the pricing of swaptions and has been
 used by several authors ever since. \shortciteN{BraceDunBarton01} among others argue that freezing the
 drift is justified, since the deviation from the original equation is small in
 several measures.

 Although the frozen drift approximation is the simplest and most popular solution,
 it is well-known that it does not yield acceptable results, especially
 for exotic derivatives and longer horizons. Therefore, several other
 approximations have been developed in the literature. We refer the
 reader to Joshi and Stacey \citeyear{JoshiStacey08} for a detailed overview of that literature, and
 for some new approximation schemes and numerical experiments.

 Although most of this literature focuses on the lognormal LIBOR market model,
 \citeANP{GlassermanMerener03} (\citeyearNP{GlassermanMerener03},
 \citeyearNP{GlassermanMerener03b}) have developed approximation schemes for the
 pricing of caps and swaptions in jump-diffusion \lib market models.

 In this article we develop a general method for the approximation of the random
 terms that enter into the drift of LIBOR models. In particular, by applying Picard
 iterations we develop a generic approximation scheme. The method we develop yields
 more accurate results than the
 frozen drift approximation, while having the added feature that the individual
 rates can be evolved independently in a Monte Carlo simulation. This enables the
 use of parallel computing in the maturity dimension. Moreover, our method is universal
 and can be applied to any \lib model driven by a general \smmg. We illustrate the
 accuracy and speed of our method in a case where LIBOR rates are driven by a
 normal inverse Gaussian process.

\section{The L\'evy LIBOR model}
\label{LevyLIBOR}

The \lev LIBOR model was developed by \citeN{EberleinOezkan05}, following the
seminal articles of \shortciteN{SandmannSondermannMiltersen95},
Miltersen et al. 
\citeyear{MiltersenSandmannSondermann97} and
\shortciteN{BraceGatarekMusiela97} on LIBOR market models driven by Brownian
motion; see also \citeN{GlassermanKou03} and \citeN{Jamshidian99} for LIBOR
models driven by jump processes and general semimartingales respectively. The
\lev \lib model is a \textit{market model} where the forward LIBOR rate is
modeled directly, and is driven by a \tih \lev process.

Let $0=T_0<T_{1}<\cdots<T_{N}<T_{N+1}=T_*$ denote a discrete tenor
structure where $\delta_i=T_{i+1}-T_{i}$, $i\in\{0,1,\dots,N\}$. Consider a
complete stochastic basis $(\Omega, \F,\mathbf{F},\P_{T_*})$ and a \tih \lev
process \prozess[H] satisfying standard assumptions such as the existence of
exponential moments and absolutely continuous characteristics. The law of $H$
is described by the \lev--Khintchine formula:
\begin{align}\label{char-fun}
\E_{\pt[*]}\!\left[\e^{iuH_{t}}\right] = \exp\bigg(\int_{0}^{t} \kappa_s(iu)\ud s\bigg).
\end{align}
Here $\kappa_s$ is the \emph{cumulant generating function} associated to
the infinitely divisible distribution with \lev triplet ($0,c,F^{T_*}$), i.e.
for $u\in\mathbb{R}$ and $s\in[0,T_*]$
\begin{align}\label{cumulant}
\kappa_s(iu) = -\frac{c_s}{2}u^{2}
             + \int_{\R}(\e^{iux}-1-iux)F_s^{T_*}(\ud x).
\end{align}
The canonical decomposition of $H$ is:
\begin{align}\label{canon-LIBOR}
H = \int_0^\cdot \sqrt{c_s}\ud W_s^{T_*}
  + \int_0^\cdot\int_{\R}x(\mu^H-\nu^{T_*})\dsdx,
\end{align}
where $W^{T_*}$ is a $\P_{T_*}$-standard Brownian motion, $\mu^H$ is the random
measure associated with the jumps of $H$ and $\nu^{T_*}$ is the
$\P_{T_*}$-compensator of $\mu^H$. We further assume that the following
conditions are in force.
\begin{description}
\item[(LR1)]
 For any maturity $T_{i}$ there exists a bounded, continuous, deterministic
 function $\lambda(\cdot,T_{i}):[0,T_i]\rightarrow \R$, which represents the
 volatility of the forward LIBOR rate process $L(\cdot, T_{i})$. Moreover, we
 assume that (i) for all $s\in[0,T_*]$, there exist $M,\epsilon>0 $ such that
 $\int_0^{T_*}\int_{\{|x|>1\}}\e^{ux} F_t(\ud x)\ud t<\infty$, for
 $u\in[-(1+\varepsilon)M,(1+\varepsilon)M]$, and (ii) for all $s<T_i$
\begin{align*}
\sum_{i=1}^N \big|\lambda(s,T_i)\big|\leq M.
\end{align*}
\item[(LR2)]
  The initial term structure $B(0,T_i)$, $1\leq i\leq N+1$, is strictly
  positive and strictly decreasing. Consequently, the initial term structure of
  forward LIBOR rates is given, for $1\leq i\leq N$, by
\begin{align}\label{i-val}
L(0,T_i)=\frac{1}{\delta_i}\left(\frac{B(0,T_i)}{B(0,T_i+\delta_i)}-1\right)>0.
\end{align}
\end{description}

The construction of the model starts by postulating that the dynamics of the forward LIBOR
rate with the longest maturity $L(\cdot,T_N)$ is driven by the \tih \lev process
$H$ and evolve as a martingale under the terminal forward measure $\P_{T_*}$.
Then, the dynamics of the LIBOR rates for the preceding maturities are
constructed by backward induction; they are driven by the same process $H$ and
evolve as martingales under their associated forward measures. For the full
mathematical construction we refer to \citeN{EberleinOezkan05}.

We will now proceed to introduce the stochastic differential equation that the
dynamics of log-LIBOR rates satisfy under the terminal measure $\P_{T_*}$. This
will be the starting point for the approximation method that will be developed
in the next section.

In the \lev LIBOR model the dynamics of the LIBOR rate $L(\cdot,\ts[i])$
under the terminal forward measure \pt[*] are given by
\begin{align}\label{LIBOR-dyn-PT}
\LibT = L(0,T_i)
 \exp\left( \int_0^t b(s,T_i)\ud s+\int_0^t \volT\ud H_s \right),
\end{align}
where \prozess[H] is the $\P_{T_*}$-\tih \lev process. The drift term $b(\cdot,\ts[i])$
is determined by no-arbitrage conditions and has the form
\begin{align}\label{LIBOR-drift-PT}
b(s,\ts[i])
 & = -\half \vol2T c_s
     - c_s\volT \sum_{l=i+1}^{N}\frac{\delta_l L(s-,T_l)}{1+\delta_l L(s-,T_l)}\lambda(s,T_l)
     \nonumber\\&\quad
     - \int_{\R}\left(\Big(\e^{\volT x}-1\Big)\prod_{l=i+1}^{N}\beta(s,x,T_l)
                      -\volT x\right)F_s^{T_*}(\ud x),
\end{align}
where
\begin{align}\label{Levy-LIBOR-beta}
\beta(t,x,T_l,)
 = \frac{\delta_l L(t-,T_l)}{1+\delta_l L(t-,T_l)}\Big(\e^{\lambda(t,T_l)x}-1\Big) +1.
\end{align}
Note that the drift term in \eqref{LIBOR-dyn-PT} is random, therefore we are dealing with a general
semimartingale, and not with a \lev process. Of course, $L(\cdot,\ts[i])$ is not
a $\P_{T_*}$-martingale, unless $i=N$ (where we use the conventions $\sum_{l=1}^0=0$
and $\prod_{l=1}^0=1$).

Let us denote by $Z$ the log-\lib rates, that is
\begin{align}\label{log-LIB-SIE}
Z(t,T_i)
 &:= \log L(t,T_i) \nonumber\\
 &= Z(0,T_i) + \int_0^t b(s,T_i)\ud s + \int_0^t \volT\ud H_s,
\end{align}
where $Z(0,T_i)=\log L(0,T_i)$ for all $i\in\{1,\dots,N\}$.

\begin{remark}
Note that the martingale part of $Z(\cdot,T_i)$, i.e. the stochastic integral
$\int_0^\cdot\volT\ud H_s$,
is a \tih \lev process. However, the random drift term destroys the \lev property
of $Z(\cdot,T_i)$, as the increments are no longer independent.
\end{remark}

\section{Picard approximation for LIBOR models}

The log-\lib can be alternatively described as a solution to the following linear SDE
\begin{align}\label{log-LIB-SDE-2}
\ud Z(t,T_i) &= b(t,T_i)\dt + \lambda(t,T_i)\ud H_t,
\end{align}
with initial condition $Z(0,T_i)=\log L(0,T_i)$.
Let us look further into the above SDE for the log-\lib rates. We introduce the term $Z(\cdot)$
in the drift term $b(\cdot,T_i;Z(\cdot))$ to make explicit that the log-\lib rates
depend on all subsequent rates on the tenor.

The idea behind the Picard approximation scheme is to approximate the drift term
in the dynamics of the \lib rates; this approximation is achieved by the Picard iterations
for \eqref{log-LIB-SDE-2}. The first Picard iteration for \eqref{log-LIB-SDE-2} is
simply the initial value, i.e.
\begin{align}
Z^{(0)}(t,T_i) = Z(0,T_i),
\end{align}
while the second Picard iteration is
\begin{align}\label{Picard-2}
Z^{(1)}(t,T_i)
 &= Z(0,T_i) + \int_0^t b(s,T_i;Z^{(0)}(s))\ds + \int_0^t\lambda(s,T_i)\ud H_s \nonumber\\
 &= Z(0,T_i) + \int_0^t b(s,T_i;Z(0))\ds + \int_0^t\lambda(s,T_i)\ud H_s.
\end{align}
Since the drift term $b(\cdot,T_i;Z(0))$ is deterministic, as the random terms
have been replaced with their initial values, we can easily deduce that the second
Picard iterate $Z^{(1)}(\cdot,T_i)$ is a \lev process.

Comparing \eqref{Picard-2} with \eqref{log-LIB-SIE} it becomes evident
that we are approximating the semimartingale $Z(\cdot,T_i)$ with the \tih \textit{\lev
process} $Z^{(1)}(\cdot,\ts[i])$.

\subsection{Application to LIBOR models}

In this section, we will apply the Picard approximation of the log-LIBOR
rates $Z(\cdot,T_i)$ by $Z^{(1)}(\cdot,T_i)$ in order to derive a \emph{strong},
i.e. pathwise, approximation for the dynamics of log-\lib rates. That is, we
replace the random terms in the drift $b(\cdot,T_i;Z(\cdot))$ by the \lev process
$Z^{(1)}(\cdot,T_i)$ instead of the semimartingale $Z(\cdot,T_i)$. Therefore,
the dynamics of the \textit{approximate} log-\lib rates are given by
\begin{align}
\widehat{Z}(t,T_i)
 &= Z(0,T_i) + \int_0^t b(s,\ts[i];Z^{(1)}(s))\ds + \int_0^t \lambda(s,T_i)\ud H_s,
\end{align}
where the drift term is provided by
\begin{align}
 \label{bdrift}
b(s,\ts[i];Z^{(1)}(s))
 & = -\half \vol2T c_s
     - c_s\volT \sum_{l=i+1}^{N}
          \frac{\delta_l \e^{Z^{(1)}(s-,T_l)}}{1+\delta_l \e^{Z^{(1)}(s-,T_l)}}\lambda(s,T_l)
     \nonumber\\&\,\,
     - \int_{\R}\left(\Big(\e^{\volT x}-1\Big)\prod_{l=i+1}^{N}\widehat\beta(s,x,T_l)
                      -\volT x\right)F_s^{T_*}(\ud x),
\end{align}
with
\begin{align}
\widehat\beta(t,x,T_l,)
 = \frac{\delta_l \exp\big(Z^{(1)}(t-,T_l)\big)}
        {1+\delta_l \exp\big(Z^{(1)}(t-,T_l)\big)}\Big(\e^{\lambda(t,T_l)x}-1\Big) +1.
\end{align}

The main advantage of the Picard approximation is that the resulting SDE
for $\widehat{Z}(\cdot,T_i)$ can be simulated more easily than the equation for
$Z(\cdot,T_i)$. Indeed, looking at \eqref{log-LIB-SDE-2} and \eqref{LIBOR-drift-PT}
again, we can observe that each LIBOR rate $L(\cdot,T_i)$ depends on all subsequent
rates $L(\cdot,T_l)$, $i+1\le l\le N$. Hence, in order to simulate $L(\cdot,T_i)$,
we should start by simulating the furthest rate in the tenor and proceed iteratively
from the end. On the contrary, the dynamics of $\widehat{Z}(\cdot,T_i)$ depend
only on the \lev processes $Z^{(1)}(\cdot,T_l)$, $i+1\le l\le N$, which are
independent of each other. Hence, we can use \textit{parallel computing} to
simulate all approximate \lib rates simultaneously. This significantly increases
the speed of the Monte Carlo simulations which will be demonstrated in the numerical example.

\subsection{Drift expansion}
\label{driftsecond}

Let us look now at the drift term in \eqref{LIBOR-drift-PT} more carefully.
Observe that there is a product of the form $\prod_{k=1}^N(1+a_k)$ appearing;
the expansion of this product has the following form
\begin{align}\label{product}
\prod_{k=1}^N(1+a_k)
 &= 1 + \sum_{k=1}^N a_k
  + \underbrace{\sum_{1\le i<j\le N}a_ia_j}_{\# \binom{N}{2}}
  + \underbrace{\sum_{1\le i<j<k\le N}a_ia_ja_k}_{\# \binom{N}{3}} \nonumber\\
 &\quad
  + \underbrace{\sum_{i\neq j\neq k\neq l}a_ia_ja_ka_l}_{\# \binom{N}{4}}
  + \dots
  + \prod_{k=1}^Na_k,
\end{align}
where the number of terms on the right hand side is $2^N$. Therefore, we need
to perform $2^N$ computations in order to calculate the drift of the \lib rates.
Since $N$ is the length of the tenor, it becomes apparent that this calculation
is feasible for a short tenor, but not for long tenors; e.g. for $N=40$ this
amounts to more than 1 trillion computations.

In order to deal with this computational problem, we will approximate the LHS
of \eqref{product} with the first or second order terms. Let us introduce the
following shorthand notation for convenience:
\begin{align}
 \lambda_l := \lambda(s,T_l)
 \quad\text{ and }\quad
 L_l := L(s,T_l).
\end{align}
We denote by $\mathbb{A}$ the part of the drift term that is stemming from the
jumps, i.e.
\begin{align}
\mathbb{A}
 &= \int_{\R}\left( \Big(\e^{\lambda_i x}-1\Big)
      \prod_{l=i+1}^{N} \left(1+\frac{\delta_l L_l}{1+\delta_l L_l}
            \Big(\e^{\lambda_lx}-1\Big)\right)-\lambda_i x\right)F_s^{T_*}(\ud x).
\end{align}
The first order approximation of the product term is
\begin{align}
\label{drifta}
\mathbb{A}'
 &= \int_{\R} \left( \Big(\e^{\lambda_i x}-1\Big)
       \left( 1 + \sum_{l=i+1}^{N}\frac{\delta_l L_l}{1+\delta_l L_l}
            \Big(\e^{\lambda_lx}-1\Big)\right)-\lambda_i x\right)F_s^{T_*}(\ud x) \nonumber\\
 &= \int_{\R} \left(\e^{\lambda_i x}-1-\lambda_i x\right)F_s^{T_*}(\ud x) \nonumber\\
 &\quad + \sum_{l=i+1}^{N}\frac{\delta_l L_l}{1+\delta_l L_l}
    \int_{\R} \Big(\e^{\lambda_i x}-1\Big)\Big(\e^{\lambda_lx}-1\Big)F_s^{T_*}(\ud x)\nonumber\\
 &= \kappa\big(\lambda_i\big)
  + \sum_{l=i+1}^{N}\frac{\delta_l L_l}{1+\delta_l L_l}
    \Big(\kappa\big(\lambda_i+\lambda_l\big)
           - \kappa\big(\lambda_i\big) - \kappa\big(\lambda_l\big)\Big),
\end{align}
and the order of the error is
\begin{align}
\mathbb{A} = \mathbb{A}' + O(\|L\|^2).
\end{align}
Similarly the second order approximation is provided by
\begin{align}
\label{drifta2}
\mathbb{A}''
 &= \kappa\big(\lambda_i\big)
  + \sum_{l=i+1}^{N}\frac{\delta_l L_l}{1+\delta_l L_l}
    \Big(\kappa\big(\lambda_i+\lambda_l\big)
           + \kappa\big(\lambda_i\big) + \kappa\big(\lambda_l\big)\Big) \nonumber\\
 &\quad + \sum_{i+1\le k<l\le N}
           \frac{\delta_l L_l}{1+\delta_l L_l}\frac{\delta_k L_k}{1+\delta_k L_k} \nonumber\\
 &\quad\times \Big( \kappa\big(\lambda_i+\lambda_l+\lambda_k\big)
        - \kappa\big(\lambda_i+\lambda_l\big)
        - \kappa\big(\lambda_i+\lambda_k\big) \nonumber\\
 &\qquad        - \kappa\big(\lambda_k+\lambda_l\big)
        + \kappa\big(\lambda_i\big)
        + \kappa\big(\lambda_l\big)
        + \kappa\big(\lambda_k\big) \Big),
\end{align}
and the order of the error is
\begin{align}
\mathbb{A} = \mathbb{A}'' + O(\|L\|^3).
\end{align}

\subsection{Caplets}

The price of a caplet with strike $K$ maturing at time $T_i$, using the
relationship between the terminal and the forward measures can be expressed as
\begin{align}\label{caplet}
\mathbb{C}_0(K,\ts[i])
 &= \delta B(0,\ts[*])\,
    \E_{\pt[*]}\Big[\prod_{l=i+1}^{N}\big(1+\delta L(T_i,T_l)\big)(L(T_i,T_i)-K)^+\Big].
\end{align}
This equation will provide the actual prices of caplets corresponding
to simulating the full SDE for the LIBOR rates. In order to calculate the
Picard approximation prices for a caplet we have to replace
$L(\cdot,T_\cdot)$ in \eqref{caplet} with $\widehat{L}(\cdot,\ts[\cdot])$.
Similarly, for the frozen drift approximation prices we must use
$\widehat L^0(\cdot,\ts[\cdot])$ instead of $L(\cdot,T_\cdot)$.

\section{Numerical illustration}
\label{numerics}

The aim of this section is to demonstrate the accuracy and efficiency of the
Picard approximation scheme for the valuation of options in the \lev \lib model
compared to the ``frozen drift'' approximation. In addition, we investigate the
accuracy of the drift expansions in section \ref{driftsecond}. We will
consider the pricing of caplets, although many other interest rate derivatives
can be considered in this framework.

We revisit the numerical example in \citeN[pp. 76-83]{Kluge05}. That is, we
consider a tenor structure $T_0=0, T_1=\frac12, T_2=1 \dots, T_{10}=5=T_*$,
constant volatilities
\begin{align*}
  \lambda(\cdot,T_1)= 0.20 \qquad
  \lambda(\cdot,T_2)= 0.19 \qquad
  \lambda(\cdot,T_3)= 0.18 \\
  \lambda(\cdot,T_4)= 0.17 \qquad
  \lambda(\cdot,T_5)= 0.16 \qquad
  \lambda(\cdot,T_6)= 0.15 \\
  \lambda(\cdot,T_7)= 0.14 \qquad
  \lambda(\cdot,T_8)= 0.13 \qquad
  \lambda(\cdot,T_9)= 0.12
\end{align*}
and the discount factors (zero coupon bond prices) as quoted on February 19,
2002; cf. Table \ref{tab:zcbprices}. The tenor length is constant and denoted
by $\delta=\frac12$.

{\renewcommand{\arraystretch}{1.16}
\begin{table}
 \begin{center}
  \begin{minipage}{11cm}\small
   \begin{tabular}{c|ccccc}
\hline
$T$      & 0.5\,Y&1\,Y&1.5\,Y&2\,Y&2.5\,Y\\
$B(0,T)$ & 0.9833630& 0.9647388& 0.9435826& 0.9228903& 0.9006922\\
\hline
$T$      &3\,Y&3.5\,Y&4\,Y&4.5\,Y&5\,Y\\
$B(0,T)$ & 0.8790279& 0.8568412& 0.8352144& 0.8133497& 0.7920573\\
 \hline
    \end{tabular}~\\[1ex]
   \end{minipage}
  \caption{Euro zero coupon bond prices on February 19, 2002.}
  \label{tab:zcbprices}
 \end{center}
\end{table}
}

The driving \lev process $H$ is a normal inverse Gaussian (NIG) process with
parameters $\alpha=\bar\delta=1.5$ and $\mu=\beta=0$. We denote by $\mu^H$ the
random measure of jumps of $H$ and by $\nu\dtdx=F(\dx)\dt$ the
$\P_{T_*}$-compensator of $\mu^H$, where $F$ is the \lev measure of the NIG
process. The necessary conditions are satisfied because $M=\alpha$, hence
$\sum_{i=1}^{9}|\lambda(\cdot,T_i)|=1.44<\alpha$
and $\lambda(\cdot,T_i)<\frac{\alpha}{2}$, for all $i\in\{1,\dots,9\}$.

The NIG \lev process is a pure-jump \lev process with canonical decomposition
$H=\int_0^\cdot\int_\R x (\mu^H-\nu)\dsdx.$
The cumulant generating function of the NIG distribution, for all $u\in\mathbb{C}$
with $|\Re u|\le\alpha$, is
\begin{align}
\kappa(u)
 &= \bar\delta\alpha-\bar\delta\sqrt{\alpha^2-u^2}.
\end{align}

\begin{figure}[ht!]
 \centering
 \includegraphics[width=6.25cm]{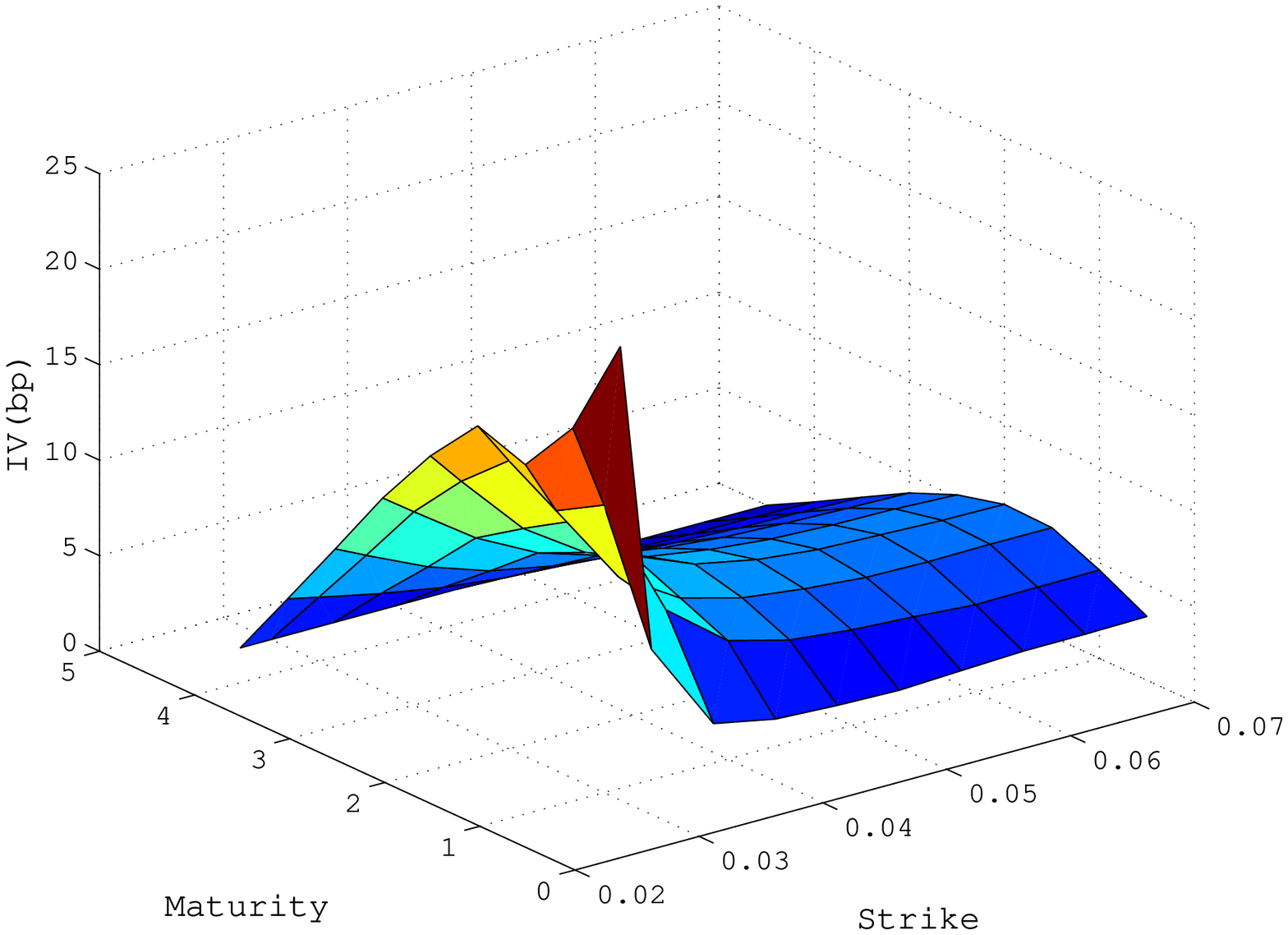}
 \includegraphics[width=6.25cm]{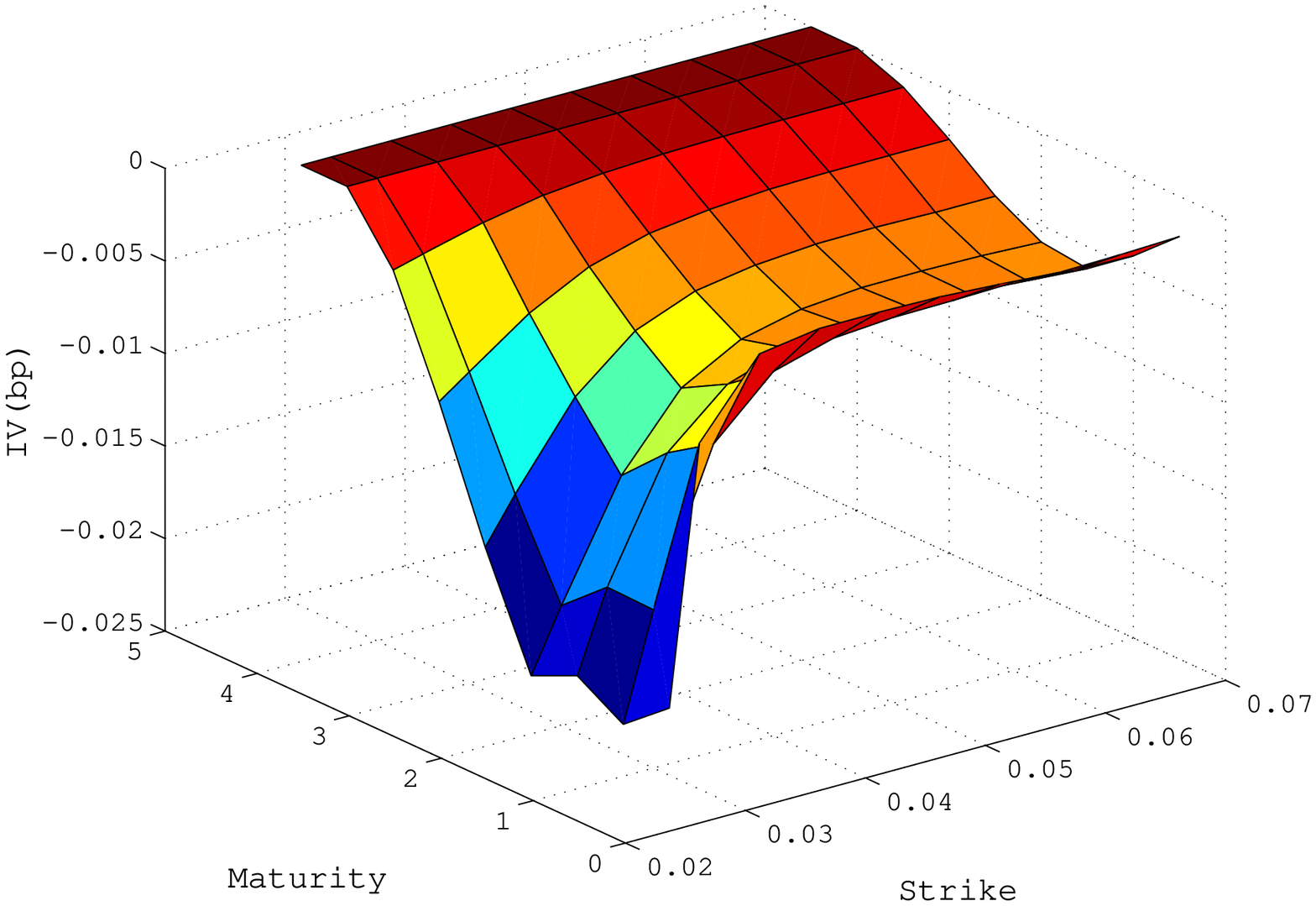}
 \caption{Difference in implied caplet volatility (in basis points) between the full SDE and the frozen
           drift prices (left), and the full SDE and the Picard approximation (right).}
 \label{fig:caplets-IV-diffs}
\end{figure}

In figure \ref{fig:caplets-IV-diffs} we plot the difference in basis points (bp)
between caplet implied volatility calculated from the full numerical solution and
implied volatilities from the frozen drift and the Picard approximation respectively.
In order to isolate the error from the two approximations we use the same discretization
grid (5 steps per tenor length) and the same pseudo random numbers (10000 paths)
in each method. The pseudo random numbers are generated from the NIG distribution
using the standard methodology described in \citeNP{Glasserman03}. The drifts
are first calculated without approximation using expression \eqref{bdrift}. It is
clear that the Picard approximation outperforms the frozen drift with an error
which is at maximum 0.023 bp. Since implied volatility is quoted in units of a
basis point then 1bp is a natural maximum tolerance level of error in an approximation.
It is clear that the error in the frozen drift approximation is significantly bigger
than one basis point throughout most strikes and maturities, with a maximum around 17bp.
Both graphs also show that the error in general increases in absolute terms the
smaller the strike.


As we established in \eqref{product} the number of terms needed to calculate the
drift grows with a rate $2^N$. In market applications $N$ is often as high as 60
reflecting a 30 year term structure with a 6 month tenor increment. At this level
even the calculation of one drift term becomes infeasible and this necessitates
the use of the approximations introduced in \eqref{drifta} and \eqref{drifta2}.
If we investigate the errors introduced by comparing them with the full numerical
solution we get an average error of 0.41 bp with max of 9.5 whereas the second
order approximation  performs much better with an average error of 0.013 and maximum
error around 0.38 bp. 

\begin{figure}[ht!]
 \centering
 \includegraphics[width=6.25cm]{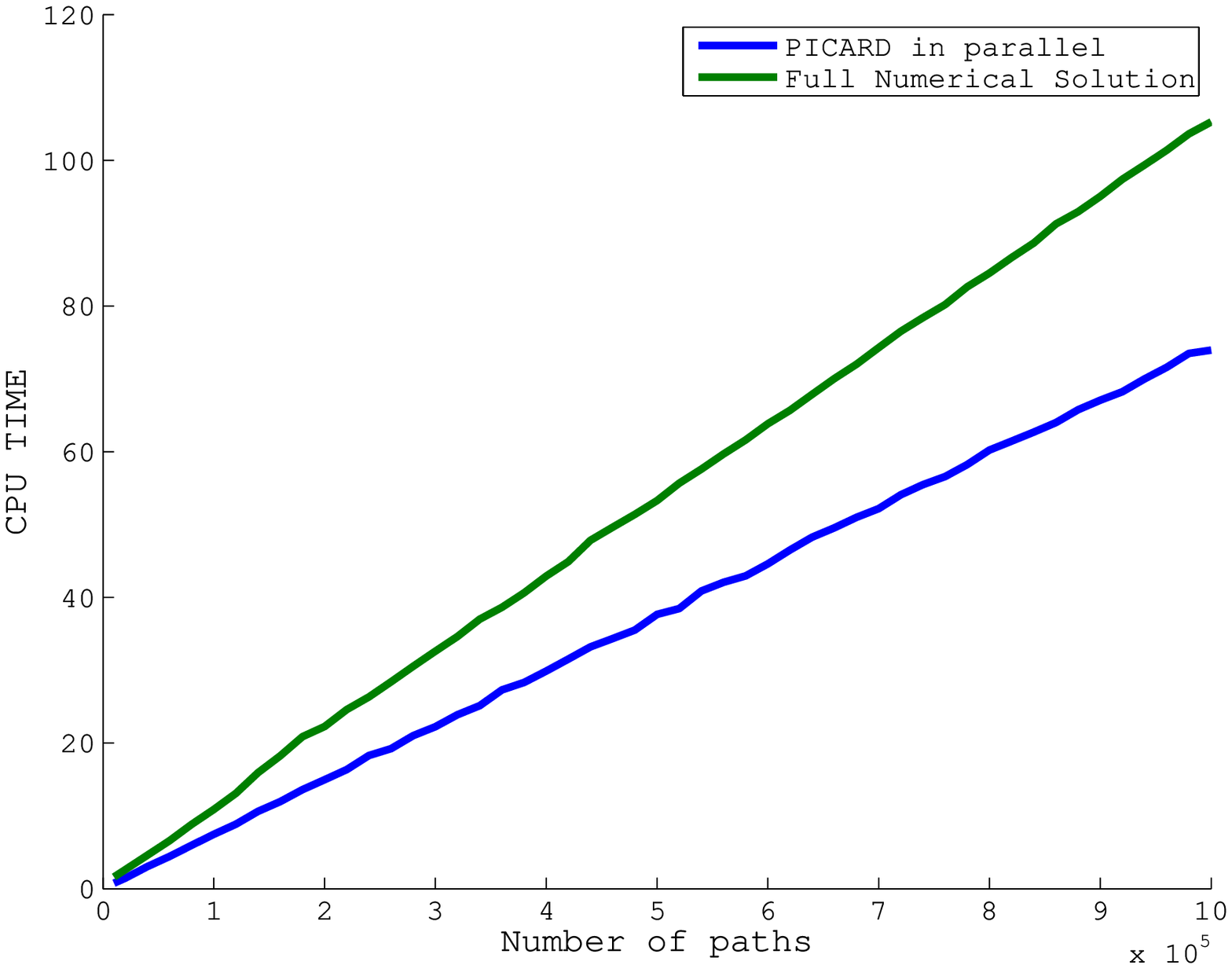}
 \includegraphics[width=6.25cm]{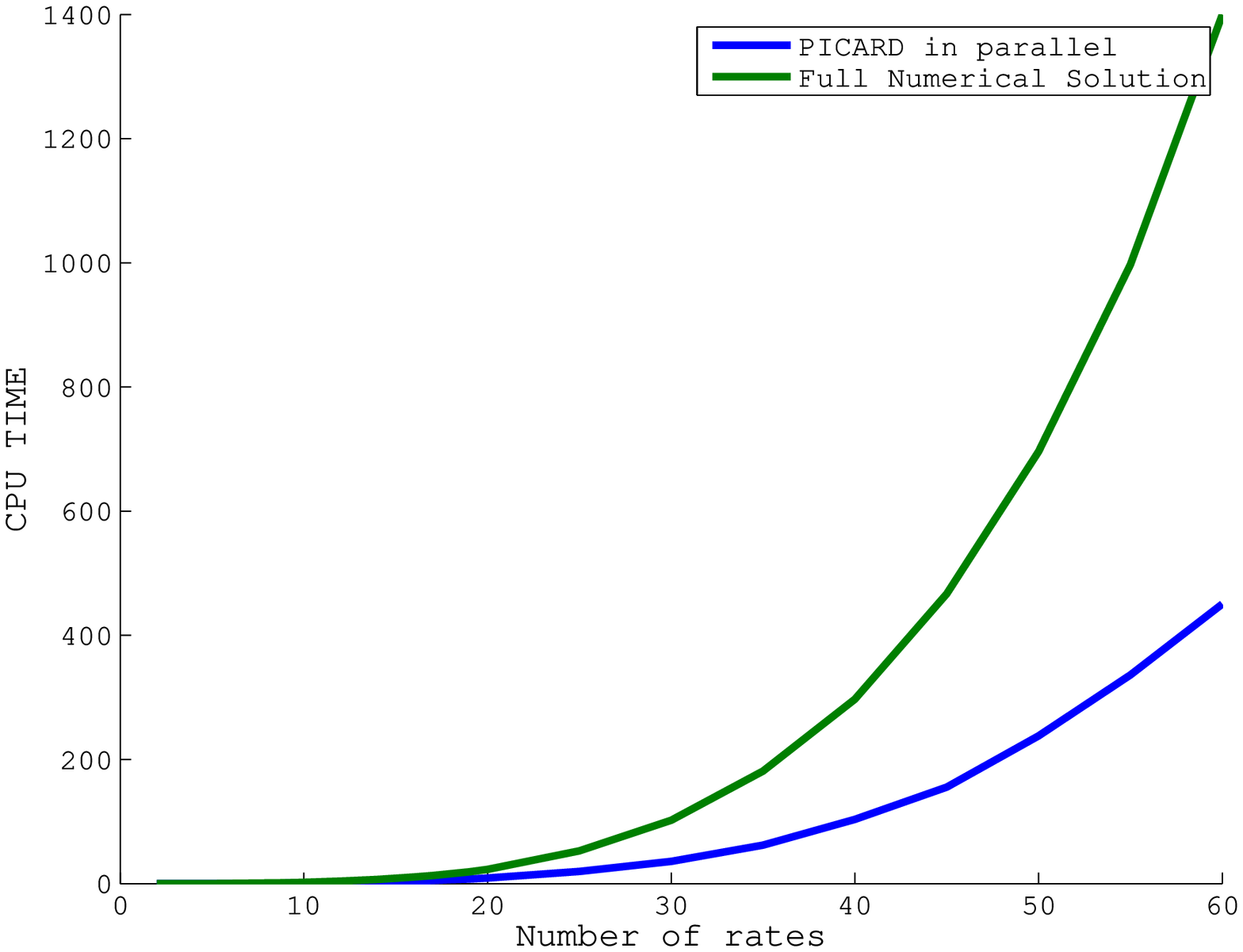}
 \caption{CPU time as function of the number of paths (left) and CPU time as a function of $N$, the number of rates (right).}
 \label{fig:caplets-CPU-TIME}
\end{figure}

In terms of computational time a large gain is realized when using the approximations
in \eqref{drifta} and \eqref{drifta2}. In the example above the CPU time for the
full numerical solution is 141 seconds but after applying the first order or second
order drift approximation it drops to 0.9 and 1.2 respectively. Adding the Picard
approximation to these three cases does not contribute to the computational speed
unless parallelization is employed. On the left in figure \ref{fig:caplets-CPU-TIME},
CPU time as function of the number of paths for the Picard approximation and the
full numerical solution is plotted. Both use the second order drift approximation
scheme in \eqref{drifta2}. The computations are done in Matlab running on an Intel
i7 processor with the capability of running 8 processes simultaneously. Here we see
the typical linear behavior as the number of paths are increased but it can be seen
that the Picard approximation has a significantly lower slope. Furthermore, on the
right when we plot CPU time as a function of rates one can see CPU time
exponentially increasing, revealing that large gains in computational time are
realizable when using the Picard approximation scheme and the drift expansion.

\section{Conclusion}

This paper derives a new approximation method for Monte Carlo derivative pricing
in LIBOR models. It is generic and can be used for any semi-martingale driven
model. It decouples the interdependence of the rates when moving them forward in
time in a simulation, meaning that the computations can be parallelized in the
maturity dimension. We have demonstrated both the accuracy and speed of the method
in a numerical example.

\bibliographystyle{chicago}
\bibliography{references}

\begin{thebibliography}{}

\bibitem[\protect\citeauthoryear{Andersen and Brotherton-Ratcliffe}{Andersen
  and Brotherton-Ratcliffe}{2005}]{AndersenBrothertonRatcliffe05}
Andersen, L. and R.~Brotherton-Ratcliffe (2005).
\newblock Extended {LIBOR} market models with stochastic volatility.
\newblock {\em J. Comput. Finance\/}~{\em 9}, 1--40.

\bibitem[\protect\citeauthoryear{Black}{Black}{1976}]{Black76}
Black, F. (1976).
\newblock The pricing of commodity contracts.
\newblock {\em J. Financ. Econ.\/}~{\em 3}, 167--179.

\bibitem[\protect\citeauthoryear{Brace, Dun, and Barton}{Brace
  et~al.}{2001}]{BraceDunBarton01}
Brace, A., T.~Dun, and G.~Barton (2001).
\newblock Towards a central interest rate model.
\newblock In E.~Jouini, J.~Cvitani{\'c}, and M.~Musiela (Eds.), {\em Option
  pricing, interest rates and risk management}, pp.\  278--313. Cambridge
  University Press.

\bibitem[\protect\citeauthoryear{Brace, G\c{a}tarek, and Musiela}{Brace
  et~al.}{1997}]{BraceGatarekMusiela97}
Brace, A., D.~G\c{a}tarek, and M.~Musiela (1997).
\newblock The market model of interest rate dynamics.
\newblock {\em Math. Finance\/}~{\em 7}, 127--155.

\bibitem[\protect\citeauthoryear{Eberlein and \"Ozkan}{Eberlein and
  \"Ozkan}{2005}]{EberleinOezkan05}
Eberlein, E. and F.~\"Ozkan (2005).
\newblock {The L\'evy LIBOR model}.
\newblock {\em Finance Stoch.\/}~{\em 9}, 327--348.

\bibitem[\protect\citeauthoryear{Glasserman}{Glasserman}{2003}]{Glasserman03}
Glasserman, P. (2003).
\newblock {\em Monte Carlo methods in financial engineering}.
\newblock Springer-Verlag.

\bibitem[\protect\citeauthoryear{Glasserman and Kou}{Glasserman and
  Kou}{2003}]{GlassermanKou03}
Glasserman, P. and S.~G. Kou (2003).
\newblock The term structure of simple forward rates with jump risk.
\newblock {\em Math. Finance\/}~{\em 13}, 383--410.

\bibitem[\protect\citeauthoryear{Glasserman and Merener}{Glasserman and
  Merener}{2003a}]{GlassermanMerener03b}
Glasserman, P. and N.~Merener (2003a).
\newblock {Cap and swaption approximations in LIBOR market models with jumps}.
\newblock {\em J. Comput. Finance\/}~{\em 7}, 1--36.

\bibitem[\protect\citeauthoryear{Glasserman and Merener}{Glasserman and
  Merener}{2003b}]{GlassermanMerener03}
Glasserman, P. and N.~Merener (2003b).
\newblock {Numerical solution of jump-diffusion LIBOR market models}.
\newblock {\em Finance Stoch.\/}~{\em 7}, 1--27.

\bibitem[\protect\citeauthoryear{Jamshidian}{Jamshidian}{1999}]{Jamshidian99}
Jamshidian, F. (1999).
\newblock {LIBOR} market model with semimartingales.
\newblock Working Paper, NetAnalytic Ltd.

\bibitem[\protect\citeauthoryear{Joshi and Stacey}{Joshi and
  Stacey}{2008}]{JoshiStacey08}
Joshi, M. and A.~Stacey (2008).
\newblock New and robust drift approximations for the {LIBOR} market model.
\newblock {\em Quant. Finance\/}~{\em 8}, 427--434.

\bibitem[\protect\citeauthoryear{Kluge}{Kluge}{2005}]{Kluge05}
Kluge, W. (2005).
\newblock {\em {Time-inhomogeneous L\'evy processes in interest rate and credit
  risk models}}.
\newblock Ph.\ D. thesis, Univ. Freiburg.

\bibitem[\protect\citeauthoryear{Miltersen, Sandmann, and Sondermann}{Miltersen
  et~al.}{1997}]{MiltersenSandmannSondermann97}
Miltersen, K.~R., K.~Sandmann, and D.~Sondermann (1997).
\newblock Closed form solutions for term structure derivatives with log-normal
  interest rates.
\newblock {\em J. Finance\/}~{\em 52}, 409--430.

\bibitem[\protect\citeauthoryear{Sandmann, Sondermann, and Miltersen}{Sandmann
  et~al.}{1995}]{SandmannSondermannMiltersen95}
Sandmann, K., D.~Sondermann, and K.~R. Miltersen (1995).
\newblock Closed form term structure derivatives in a {Heath--Jarrow--Morton}
  model with log-normal annually compounded interest rates.
\newblock In {\em Proceedings of the Seventh Annual European Futures Research
  Symposium Bonn}, pp.\  145--165.
\newblock Chicago Board of Trade.

\end{thebibliography}

\end{document}